\documentclass[pre,showpacs,preprintnumbers,amsmath,amssymb]{revtex4}
\usepackage{amssymb,epsf}
\usepackage{latexsym}
\usepackage{color}

\begin{document}

\title{Relativistic three-partite non-locality}
\author{Hooman Moradpour$^{1}$\footnote{h.moradpour@riaam.ac.ir} and
Afshin Montakhab$^2$\footnote{montakhab@shirazu.ac.ir}}
\address{$^1$ Research Institute for Astronomy and Astrophysics of Maragha
(RIAAM), P.O. Box 55134-441, Maragha, Iran,\\
$^2$ Department of Physics, College of Sciences, Shiraz
University, Shiraz 71946-84795, Iran.}
\date{\today}
\begin{abstract}
Bell-like inequalities have been used in order to distinguish
non-local quantum pure states by various authors. The behavior of
such inequalities under Lorentz transformation has been a source of
debate and controversies in the past. In this paper, we consider the
two most commonly studied three-particle pure states, that of W and
GHZ states which exhibit distinctly different type of entanglement.
We discuss the various types of three-particle inequalities used in
previous studies and point to their corresponding shortcomings and
strengths. Our main result is that if one uses Czachor's
relativistic spin operator and Svetlichny's inequality as the main
measure of non-locality and uses the same angles in the rest frame
($S$) as well as the moving frame ($S^{\prime}$), then maximally
violated inequality in $S$ will decrease in the moving frame, and
will eventually lead to lack of non-locality ( i.e. satisfaction of
inequality) in the $v \rightarrow c$ limit. This is shown for both
the GHZ and W states and in two different configurations which are
commonly studied (\textbf{Case $I$} and \textbf{Case $II$}). Our
results are in line with a more familiar case of two particle case.
We also show that the satisfaction of Svetlichny's inequality in the
$v\rightarrow c$ limit is independent of initial particles'
velocity. Our study shows that whenever we use Czachor's
relativistic spin operator, results draws a clear picture of
three-particle non-locality making its general properties consistent
with previous studies on two-particle systems regardless of the W
state or the GHZ state is involved. Throughout the paper, we also
address the results of using Pauli's operator in investigating the
behavior of $|S_v|$ under LT for both of the GHZ and W states and
two cases (\textbf{Case $I$} and \textbf{Case $II$}). Our
investigation shows that the violation of $|S_v|$ in moving frame
depends on the particle's energy in the lab frame, which is in
agreement with some previous works on two and three-particle
systems. Our work may also help us to classify the results of using
Czachor's and Pauli's operators to describe the spin entanglement
and thus the system spin in relativistic information theory.
\end{abstract}
 \maketitle
\bigskip
\section{Introduction \label{Introduction}}
The non-local feature of the Quantum Mechanics was first pointed out
in a paper by Einstein, Podolsky and Rosen, otherwise known as EPR
\cite{EPR}. This non-locality may yield to the entanglement
phenomenon which restrains decomposing quantum state of a system
into a product state including the states of its basic constituents
\cite{BA}. Entangled states may violate Bell's inequality, which
provides a theoretical basis for investigating entanglement in a
two-partite system \cite{Bell,CHSH}. It was Aspect and coworkers who
first verified such non-local behavior in their experiments
\cite{aspect1,aspect2,aspect3}. Entanglement can be related to the
concept of information via entropy which has attracted a large
amount of attention in recent years due to its applications in
quantum computation, teleportation and cryptography, among others
\cite{aud,Nilsen}.

Meanwhile, relativistic considerations attracted various authors
early on. On one hand, one expects the amount of information to be
independent of inertial observer \cite{terno}, while on the other
hand, probability distributions may depend on the frame of reference
\cite{wald}. Peres and co-workers have considered a single free
spin-$\frac{1}{2}$ particle and shown that the reduced density
matrix for its spin is not covariant under Lorentz transformation
(LT). Briefly, they proved that the spin entropy depends on the
frame due to Wigner rotation \cite{peres1}. Essentially, it was
proved that the single particle system can possess non-locality
\cite{vedral1,vedral2} and such non-locality changes under (LT)
\cite{vedral3}.

Perhaps more importantly, many authors have considered relativistic
effects on bi-partite entanglement in a two-particle systems
\cite{alsing,gingrich,li,rem,caban,pacho,ahn,terashima,terashima1,lee,kim,jordan,cafaro,ref1}.
Typically one is concerned with how entanglement and/or non-locality
is related in a lab frame ($S$) to a moving frame ($S^{\prime}$)
under LT. Authors in \cite{gingrich,li,rem,caban,pacho,ahn} have
considered a situation where Bell's inequality is maximally violated
in $S$. Keeping the same setup of angles for Bell's operator, they
find that this function is a decreasing function of the boost
velocity. Furthermore, as the speed of the boost approaches the
speed of light, they find that Bell's inequality is satisfied
indicating lack of non-locality in the $v\rightarrow c$ limit.
However, authors in \cite{alsing,terashima,terashima1,lee,kim} have
shown that if one Lorentz transforms the quantum states as well as
spin operators, then the maximal violation of Bell's inequality in
$S$ remains intact in the $S^{\prime}$. However, more recently, some
authors have shown that under certain LT, total entanglement
(entropy) of the system depends on the frame \cite{jordan,cafaro}.
Also, similar results have been shown in curved spacetime
\cite{ref1,fu,al,ma,le}. It has also been shown that acceleration
effects do not preserve the amount of entanglement
\cite{tera,shi,ball,ver,fun}. Therefore, the invariance of
bi-partite entanglement seems somewhat suspect at the moment.

The case of a three-particle system offers interesting
possibilities. On the one hand, it is a natural extension of
one-particle and two-particle studies mentioned above. On the other
hand, it opens the possibility of studying different types of
(multi-partite) entanglement included in such systems, but not
present in one or two-particle systems. It is well-known that the
behavior of bi-partite entanglement is different from multi-partite
entanglement \cite{Mitchel,montakhab,montakhabamico}. The most
commonly studied three-particle states are the W state
\begin{eqnarray}
|W\rangle &=&
\frac{1}{\sqrt{3}}(|++-\rangle+|+-+\rangle+|-++\rangle),
\end{eqnarray}
and the GHZ state
\begin{eqnarray}
|GHZ\rangle &=& \frac{1}{\sqrt{2}}(|+++\rangle +|---\rangle),
\end{eqnarray}
which contain genuine three-partite entanglement \cite{GHZ}. In the
above equation, $+$ and $-$ denote up and down components of the
spin along the $z$ direction, respectively.

Extending Bell's inequalities to a three-particle case was first
done by Svetlichny $(|S_v|)$ \cite{svet}. Shortly after, Mermin
$(|M|)$ also offered another such generalization \cite{mermin}.
Later on, Collins $(|M^{\prime}|)$ et al. also offered another
generalization \cite{cal,cerc}. They appear as follows:
\begin{eqnarray}\label{svet}
|S_v|&=&|E(ABC)+E(ABC')+E(AB'C)+E(A'BC) \nonumber \\
&-&E(A'B'C')-E(A'B'C)-E(A'BC')-E(AB'C')|\leq 4,\nonumber\\
|M|&=&|E(ABC')+E(AB'C)+E(A'BC)-E(A'B'C')|\leq 2,\ \nonumber \\
|M^{\prime}|&=&|E(ABC)-E(A'B'C)-E(A'BC')-E(AB'C')|\leq 2.
\end{eqnarray}
In the above equations, $A$ and $A'$ are possible measurements on
particle $1$. Same relations are valid for particles $2$ and $3$
with possible measurements $B$, $B'$ and $C$, $C'$, respectively.
$E(ABC)$ represents the expectation value of the product measurement
outcomes of the observable $A$, $B$ and $C$. For example,
\begin{eqnarray}\label{cf}
E_{GHZ}(ABC)&=& \langle GHZ|\sigma(\widehat{n_1})\otimes
\sigma(\widehat{n_2}) \otimes \sigma(\widehat{n_3})|GHZ\rangle =
\cos(\phi_1+\phi_2+\phi_3),
\end{eqnarray}
where $\sigma(\widehat{n_i})$ is the Pauli spin operator of the
i$^{\textmd{th}}$ particle in the $xy$ plane and $\phi_i$ is the
azimuthal angle of the vector $\widehat{n_i}$. It has been shown
that maximal violation of $|S_v|=4\sqrt{2}$ occurs for GHZ state for
angles which satisfy $\Sigma \phi_i = (m+\frac{3}{4})\pi$ and
$\Sigma \phi'_i = (m+\frac{9}{4})\pi$, with $(m=0,\pm1,\pm2,...)$
\cite{cerc}. For W state we restrict our measurement in the $xz$
plane where
\begin{eqnarray}\label{cf2}
&E_{W}(ABC)= \langle W|\sigma(\widehat{n_1})\otimes
\sigma(\widehat{n_2})\otimes
\sigma(\widehat{n_3})|W\rangle = \nonumber \\
&-\frac{2}{3}\cos(\theta_1+\theta_2+\theta_3)-\frac{1}{3}\cos\theta_1
\cos\theta_2 \cos\theta_3,
\end{eqnarray}
where $\theta_i$ specifies the polar angle measurement direction of
the i$^{\textmd{th}}$ spin observable. For $W$ state maximal
violation of $S_v$ has been shown to be $|S_v|=4.354$ which occur
for $\theta_i=35.264^{\circ}$ and
$\theta'_i=\pi-\theta_i=144.736^{\circ}$, $\forall \ i$ \cite{cerc}.
Additionally, the maximal values of $|M|$ (as well as
$|M^{\prime}|$) have been shown to be equal to $|M|=|M^{\prime}|=4$
and $|M|=|M^{\prime}|=3.046$ for $GHZ$ and $W$ states, respectively
\cite{cerc}.

However, there are some apparent inconsistencies between these three
measures of non-locality. One can show that when $S_v$ is maximally
violated, then $|M| =|M^{\prime}|=\frac{|S_v|}{2}$ which indicates
that $|M|$ and $|M^{\prime}|$ will not be maximally violated
\cite{cerc}. In other words, no given set of angles will cause the
three measures to obtain their maximum value. Furthermore, one can
show that $|M|$ and $|M^{\prime}|$ can show opposite behavior. For
example, when $|M|$ is maximally violated, $|M^{\prime}|$ is
satisfied and vice versa \cite{cerc}. Roy has shown that the upper
bound of the Mermin's inequality should be corrected \cite{roy}, and
it was proven that the three-particle Bell's like inequalities, such
as $|M|$ and $|M^{\prime}|$, which include four of the correlation
functions $(E(ABC))$ can be violated by a hybrid local-nonlocal
hidden variables model \cite{cerc,cal}. Since states which include
genuine three-partite non-locality can only violate the $|S_v|$
inequality \cite{svet,cerc}, this inequality can be used to
distinguish such states (such as the W and GHZ states) from other
states which does not include genuine three-partite non-locality
(such as three-particle systems with bi-partite non-locality). It is
also believed that genuine three-partite entanglement is observed
when either $|M|$ or $|M^{\prime}|$ is violated by greater amount
than $2\sqrt{2}$ \cite{gisin}, which is less than the value one gets
for the W state ($|M|=|M^{\prime}|=3.046$) a clear contradiction
\cite{cerc}. However, it was proved that $|S_v|$ can distinguish
between the GHZ and the W states in a three-particle non-local
system when its value possess a certain well-defined value of
$|S_v|=4.354$ which is, in fact, its maximum value for W state
\cite{Mitchel}. Indeed, only the GHZ state can violate $|S_v|$ to
its maximum possible value $(4\sqrt{2})$ while the W state cannot
violate $|S_v|$ to its maximum possible value \cite{cerc,Mitchel}.
Therefore, because of above-mentioned shortcomings of $|M|$ and
$|M^{\prime}|$, it seems reasonable to use $|S_v|$ in order to study
the GHZ and W states \cite{cerc,Mitchel,rev}.

We should note that the correlation functions $E(ABC)$, used in
order to evaluate $|S_v|$ (as well as $|M|$ and $|M^{\prime}|$), are
stronger than required to study general three-partite non-locality
\cite{AA,rev}. This is a weakness for $|S_v|$, and yields violation
of no-signalling constraint which leads to grandfather-type
paradoxes \cite{gft}. This shortcoming of $|S_v|$ can be eliminated
by either using bi-partite correlation functions which satisfy
no-signalling constraint or time-ordered correlations
\cite{rev,ac,tom}.

Author in \cite{moradi1} has used the $|M|$ inequality and studied
the relativistic behavior of the entanglement of GHZ state, when
moving observer $(S')$ uses measurements angles same as lab frame
$(S)$. He has used Czachor's relativistic spin operator
\cite{czachor} and considered a special set of measurement apparatus
that violates the $|M|$ inequality to its maximum possible value in
the rest frame. He concludes that the violation of the $|M|$
inequality in the moving frame, when the speed of the boost
increases $(\beta \longrightarrow 1)$ and if the directions of the
measurements are fixed, depends on the energy of the particles in
the $S$ frame. This results disagrees with the two-particle
non-local systems
\cite{alsing,gingrich,li,rem,caban,pacho,ahn,terashima,terashima1,lee,kim}.
Similar to the Bell state
\cite{alsing,terashima,terashima1,lee,kim}, Moradi et al. have shown
that in the moving frame, by choosing special measurements, one can
find maximum violation of the $|M|$ inequality \cite{moradi2}.
Finally, author compared his results with those of attempt in which
two-particle pure entangled states (the Bell states) are studied
\cite{ahn}, and concludes that the behaviors of non-locality stored
in the Bell and GHZ states under LT differ from each other if the
moving and lab frames use Czachor's relativistic spin operator and
the same special set of measurements violating the corresponding
Bell-like inequality to its maximum violation amount in the lab
frame. Similar results where authors have used Pauli operators as
the relativistic spin operator instead of the Czachor's relativistic
spin operator, can be found in \cite{you}. Briefly, if the moving
frame uses the same measurements as the lab frame and the Mermin's
inequality as a witness of the non-locality in the GHZ state, then
the violation of the inequality in the $S^{\prime}$ will depend on
the energy of the particles in the $S$ frame as the velocity of the
boost reaches that of light \cite{moradi1}. Since the entanglement
of the W state differs from that of the GHZ state, results obtained
in \cite{moradi1,you} cannot be generalized to the W state, and
therefore, the behavior of entanglement stored in the W state under
LT is completely unknown.

Moreover, if one compares the results of You et al. \cite{you} with
those of two-particle studies, obtained in
\cite{terashima,terashima1}, he finds that the behavior of
bi-partite non-locality under LT is similar to the behavior of
non-locality stored in the GHZ under LT if the lab and moving frames
use Pauli's spin operator and the same set of measurements violating
the corresponding Bell-like inequality to its maximum violation
amount in the lab frame. In this situation, the corresponding
Bell-like inequality is violated to its maximum violation amount in
the $\beta\rightarrow1$ limit for the low energy particles, and it
is not violated for the high energy particles in the
$\beta\rightarrow1$ limit. Here, we should note that the results of
studying the behavior of the Bell state under LT by considering
Pauli's spin operator and the special set of measurements violating
the Bell's inequality to its maximum violation amount in the lab
frame \cite{terashima,terashima1} differ from those of in which
Czachor's relativistic spin operator is considered instead of the
Pauli operator \cite{ahn}. Indeed, in the moving frame and
independently of particles energy, Bell's inequality is preserved in
the $\beta\rightarrow1$ limit if one uses Czachor's relativistic
spin operator and the special set of measurements violating the
Bell's inequality to its maximum violation amount in the lab frame
\cite{ahn}. It is also interesting to note again that although
authors in refs.~\cite{moradi1,you} use different spin operators,
their results are compatible with each other. The latter consistency
comes from considering the $|M|$ inequality and the same set of
measurements which violate the $|M|$ inequality to its maximum
violation amount in the lab frame, by authors, to study the behavior
of the GHZ state under LT \cite{mmm}. Therefore, there is an
inconsistency between the generalization of two-particle studies to
the three-particle studies whenever we compare the results of
considering Czachor's relativistic spin operator with those obtained
by considering Pauli's operator \cite{mmm}. In fact, it is shown
that if one uses Pauli's operator and the special set of
measurements violating $|S_v|$ to its maximum violation amount in
the lab frame, then the violation amount of three-particle Bell-like
inequalities depends on the particles energy \cite{mmm}. This latter
point helps us to better understand the mentioned inconsistency.
However, as we will show in the following, this inconsistency may be
completely solved if one considers the GHZ and W states and uses
Czachor's operator to investigate the behavior of $|S_v|$, $|M|$ and
$|M^{\prime}|$ under LT.

Our goal in this paper is to study the behavior of non-locality
stored in the GHZ and W states under LT by using the $|S_v|$
inequality, whiles the lab and moving frames use the same set of
measurements that violate the $|S_v|$ inequality to its maximum
possible violation amount in the lab frame. We also resolve the
above mentioned inconsistency. Since some authors \cite{you,moradi1}
have used $|M|$ in order to study the effects of LT on the GHZ
state, we also consider the $|M|$ and $|M^{\prime}|$ inequalities in
order to make a comparison with their works. In addition, this
analysis helps in clarifying previously discussed sensitivity of
three-partite Bell-like inequalities regarding measurement
directions \cite{cerc}. In order to achieve this, in section
$\textmd{II}$, we use Czachor's spin operator to calculate $|S_v|$
(as well as $|M|$ and $|M^{\prime}|$) for two typically studied
scenarios for both GHZ and W states in the $v\rightarrow c$ limit.
The results of using Pauli's operator in investigating the behavior
of $|S_v|$ under LT are also addressed throughout the paper. We
devote section $\textmd{III}$ to a summary of our results and some
concluding remarks.
\section{Three-partite non-local system under LT}
In the $S$ frame, consider a spin-$\frac{1}{2}$ particle with a
momentum vector $\overrightarrow{p}$ and a spin state
$\vert\psi\rangle=\lambda\vert+\rangle+\kappa\vert-\rangle$. The
total state of the particle can be written as:
\begin{eqnarray}
\vert \xi\rangle=\vert \overrightarrow{p}\rangle\vert\psi\rangle.
\end{eqnarray}
In the moving frame $(S')$, the state of the system is:
\begin{eqnarray}
\vert \xi' \rangle=\vert \Lambda \overrightarrow{p} \rangle
D(W(\Lambda,p)) \vert\psi\rangle.
\end{eqnarray}
In the above equation, $\Lambda \overrightarrow{p}$ denotes the
momentum of the particle in the boosted frame and $D(W(\Lambda,p))$
is the Wigner representation of the Lorentz group for the
spin-$\frac{1}{2}$ particle \cite{wigner}:
\begin{eqnarray}\label{r1}
D(W(\Lambda,p))=\cos\frac{\Omega_p}{2}+i(\overrightarrow{\sigma}.\overrightarrow{n})\sin\frac{\Omega_p}{2},
\end{eqnarray}
where $\sigma$ and $\Omega_p$ are the Pauli matrix and Wigner angle,
respectively. Also, we have:
\begin{eqnarray}
\overrightarrow{n}=\widehat{e}\times\widehat{p}.
\end{eqnarray}
where $\widehat{e}$ denotes unit vector in the boost direction.
$\widehat{p}$ is the unit vector along the momentum direction of the
particle in the $S$ frame. We consider the case in which the boost
speed is along the $\widehat{x}$ direction
$(\overrightarrow{\beta}=\beta\widehat{x})$ and the particle moves
along the $z$ direction $(\overrightarrow{p}=p_0 \widehat{z})$. In
this case we have:
\begin{eqnarray}\label{wr}
D(W(\Lambda,p))=\cos\frac{\Omega_p}{2}-i\sigma_y
\sin\frac{\Omega_p}{2},
\end{eqnarray}
where
\begin{eqnarray}\label{angle}
\tan\Omega_p=\frac{\sinh\alpha  \sinh\delta}{\cosh\alpha +
\cosh\delta}.
\end{eqnarray}
Here, $\cosh\delta=\frac{p_0}{m}$ and
$\cosh\alpha=\sqrt{1-\beta^2}$. Therefore, we have:
\begin{eqnarray}\label{rs}
D(W(\Lambda,p)) \vert + \rangle &=&
\cos\frac{\Omega_p}{2}\vert+\rangle +
\sin\frac{\Omega_p}{2}\vert-\rangle \nonumber \\
D(W(\Lambda,p)) \vert - \rangle
&=&-\sin\frac{\Omega_p}{2}\vert+\rangle +
\cos\frac{\Omega_p}{2}\vert-\rangle.
\end{eqnarray}
Also, the relativistic spin operator $\widehat{A}$ is given by
\cite{czachor}:
\begin{eqnarray}\label{rso}
\widehat{A}=\frac{(\sqrt{1-\beta^2}\overrightarrow{A}_{\bot}
+\overrightarrow{A}_{\|})\cdot\overrightarrow{\sigma}}{\sqrt{1+\beta^2[(\widehat{e}
\cdot\overrightarrow{A})^2-1]}}.
\end{eqnarray}
The subscripts $\bot$ and $\|$ denote the perpendicular and the
parallel components of the vector $\overrightarrow{A}$ to the boost
direction. In the $S$ frame, consider a three-particle system as:
\begin{eqnarray}
|\psi \rangle=\prod_i |p_i\rangle |GHZ\rangle,
\end{eqnarray}
where $p_i$ represents the $4$-momentum of the i$^{\textmd{th}}$
particle in the laboratory frame. Inserting the special set of the
angles $\phi_i=\frac{\pi}{4}$ and $\phi'_i=\frac{3\pi}{4}$ into the
Eq.~(\ref{cf}) and evaluating Eq.~(\ref{svet}), we obtain the
maximum violation $(4\sqrt{2})$ for $|S_v|$. Also for this set of
measurements we have $|M^{\prime}|=|M^{\prime}|=2\sqrt{2}$, which
according to \cite{gisin} indicates that the system only includes
bi-partite entanglement, which is clearly not the case for the GHZ
state at hand here. This confirms our motif that $|S_v|$ is a more
appropriate measure of three-particle non-locality than $|M|$ and
$|M^{\prime}|$.

We now set out to consider how $|S_v|$ behaves under LT when the
measurement setup are chosen to maximize its value in the rest frame
($S$) and keeping the condition that the same measurement setup is
used in $S^{\prime}$. We will see that under such conditions the
various three-particle systems considered will show consistent and
reasonable behavior, devoid of previous inconsistency when $|M|$
and/or $|M^{\prime}|$ are used to evaluate non-locality under LT.
Our results are also in line with well-known results of two-particle
systems which have been studied using Bell's inequality
\cite{gingrich,li,rem,caban,pacho,ahn,terashima,terashima1,lee,kim}.

\underline{\textbf{Case $I$}}: We first consider the situation where
all particles move along the $z$ axis with the same momenta, in the
$S$ frame, i.e. $\overrightarrow{p}_i=p_0\widehat{z}$. In order to
satisfy the no-signalling constraint, we consider a situation where
the measurement apparatus of each particle is located with arbitrary
non-zero distance from each other along the $z$ axis. In fact, their
distances from each other should be spacelike during the
measurements \cite{nosig}. The Wigner rotation is the same as the
Eq.~(\ref{wr}). Using Eq.~(\ref{rso}) we get:
\begin{eqnarray}\label{above}
&E(ABC)=\langle GHZ^{\Lambda}|\widehat{A}\otimes \widehat{B}\otimes
\widehat{C}|GHZ^{\Lambda}\rangle=
\prod_{i=A}^C\frac{1}{\sqrt{1-\beta^2 \sin^2\phi_i}}\nonumber \\
&[\cos^3\Omega_p \cos\phi_A \cos\phi_B
\cos\phi_C-\frac{\cos\Omega_p} {\gamma^2}(\cos\phi_A \sin\phi_B
\sin\phi_C +\\ \nonumber
 &\cos\phi_C \sin\phi_B \sin\phi_A + \cos\phi_B
\sin\phi_C \sin\phi_A)].
\end{eqnarray}
where
\begin{eqnarray}
|GHZ^{\Lambda}\rangle&=&D(W(\Lambda,p))|GHZ\rangle=(\cos(\frac{\Omega_p}{2}))^3\vert
GHZ\rangle\\
\nonumber &+&\frac{(\sin(\frac{\Omega_p}{2}))^3}{\sqrt{2}}(\vert
---\rangle - \vert +++\rangle)
\\
&+&\sqrt{\frac{3}{2}}\sin(\frac{\Omega_p}{2})\cos(\frac{\Omega_p}{2})
[(\sin(\frac{\Omega_p}{2})+\cos(\frac{\Omega_p}{2}))\vert
W\rangle\nonumber \\
&+&\frac{\sin(\frac{\Omega_p}{2})-\cos(\frac{\Omega_p}{2})}{\sqrt{3}}(\vert
--+\rangle +\vert -+-\rangle + \vert +--\rangle)],\nonumber
\end{eqnarray}
is the spin state in the $S'$ frame and
$\gamma=\frac{1}{\sqrt{1-\beta^2}}$. Inserting
$\phi_i=\frac{\pi}{4}$ and $\phi'_i=\frac{3\pi}{4}$ into
Eq.~(\ref{above}), we find
\begin{eqnarray}\label{ghzcr}
E(ABC)&=&\frac{cos\Omega_p}{\sqrt{2-\beta^2}^3}((cos\Omega_p)^2-3(1-\beta^2))=-E(A'B'C')\ \nonumber\\
E(A'BC')&=&E(AB'C')=E(A'B'C)=
\frac{cos\Omega_p}{\sqrt{2-\beta^2}^3}((cos\Omega_p)^2+1-\beta^2)\ \nonumber \\
E(A'BC)&=&E(AB'C)=E(ABC')=-E(A'B'C).
\end{eqnarray}
We therefore obtain
\begin{eqnarray}\label{seven}
|M|=|M^{\prime}|=\frac{|S_v|}{2}=\frac{2|\cos\Omega_p
|}{\sqrt{2-\beta^2}^3}(\cos^2\Omega_p +3(1-\beta^2)).
\end{eqnarray}
As a check, Eq.~(\ref{seven}) reduces to the $S$ frame result
($4\sqrt{2}$) in the appropriate limit of $\beta\rightarrow0$ and
$\Omega_p\rightarrow0$. In the ultra-relativistic limit
$(\beta\rightarrow1)$ and independently of the particles' initial
energy, unlike \cite{moradi1,you,mmm}, all of the above inequalities
are satisfied, indicating the familiar result that Bell's
inequalities are satisfied in $S^{\prime}$ as $\beta\rightarrow1$
\cite{ahn}.

Now, let us point to the behavior of $|S_v|$ when Pauli's operator
is considered instead of Czachor's. It has been shown that
\cite{mmm}
\begin{eqnarray}\label{vg}
|S_v|=
4\sqrt{2}\vert(\cos(\frac{\Omega_p}{2}))^6-(\sin(\frac{\Omega_p}{2}))^6\vert,
\end{eqnarray}
which leads to
\begin{eqnarray}\label{lvg}
|S_v|\sim \sqrt{2}(\frac{1+3\Gamma^2}{\Gamma^3}),
\end{eqnarray}
in the $\beta\rightarrow1$ limit. Here,
$\Gamma=\frac{1}{\sqrt{1-\frac{v^2_0}{c^2}}}$ and $v_0$ are the
energy factor and velocity of particles in the $S$ frame,
respectively. This result shows that the inequality is preserved by
the high energy particles ($\Gamma\rightarrow\infty$) and it is
violated to the same value as the lab frame ($4\sqrt{2}$) by the low
energy particles ($\Gamma\rightarrow1$) \cite{mmm}. Such results are
clearly in contrast to the case of Czachor's operator.

By following the above recipe for the W state we get
\begin{eqnarray}
\vert W \rangle^\Lambda
&=&\sqrt{3}\sin(\frac{\Omega_p}{2})\cos(\frac{\Omega_p}{2})[-\cos(\frac{\Omega_p}{2})\vert+++\rangle
+\sin(\frac{\Omega_p}{2}) \vert---\rangle]\nonumber
\\&+&[(\cos(\frac{\Omega_p}{2}))^3-2\cos(\frac{\Omega_p}{2})
(\sin(\frac{\Omega_p}{2}))^2]\vert W\rangle\nonumber
\\&+&\frac{[2\sin(\frac{\Omega_p}{2})
(\cos(\frac{\Omega_p}{2}))^2-(\sin(\frac{\Omega_p}{2}))^3]}{\sqrt{3}}(\vert
--+\rangle +\vert -+-\rangle + \vert +--\rangle)
\end{eqnarray}
for the spin state in the moving frame, and
\begin{eqnarray}
&E&_W(\theta_1\theta_2\theta_3)=\frac{1}{3}(A_{11}(\theta_1)A_{22}(\theta_2)A_{11}(\theta_3)+
A_{11}(\theta_1)A_{11}(\theta_2)A_{22}(\theta_3)\\ \nonumber
&+&A_{22}(\theta_1)A_{11}(\theta_2)A_{11}(\theta_3)
+A_{11}(\theta_1)A_{12}(\theta_2)A_{21}(\theta_3)+A_{11}(\theta_1)A_{21}(\theta_2)A_{12}(\theta_3)\\
\nonumber
&+&A_{11}(\theta_2)A_{12}(\theta_1)A_{21}(\theta_3)+A_{11}(\theta_2)A_{21}(\theta_1)A_{12}(\theta_3)+A_{11}(\theta_3)A_{12}(\theta_1)A_{21}(\theta_2)
\\ \nonumber
&+&A_{11}(\theta_3)A_{21}(\theta_1)A_{12}(\theta_2)),
\end{eqnarray}
where we have
\begin{eqnarray}
&A_{11}(\theta)=-A_{22}(\theta)=\frac{1}{\sqrt{1-\beta^2
\sin^2\theta}}
(\frac{\cos\theta\cos\Omega_p}{\gamma}+\sin\theta\sin\Omega_p),\\
\nonumber &A_{12}(\theta)=A_{21}(\theta)=\frac{1}{\sqrt{1-\beta^2
\sin^2\theta}}
(-\frac{\cos\theta\sin\Omega_p}{\gamma}+\sin\theta\cos\Omega_p).
\end{eqnarray}
for the correlation function $E_W(\theta_1\theta_2\theta_3)$. In the
$\gamma=1, \Omega_p=0$ limit ($S$ frame) we have:
\begin{eqnarray}
&A_{11}(\theta)=-A_{22}(\theta)=\cos\theta \\ \nonumber
&A_{12}(\theta)=A_{21}(\theta)=\sin\theta,
\end{eqnarray}
which leads to
\begin{eqnarray}
E_W(\theta_1\theta_2\theta_3)&=&-\cos\theta_1\cos\theta_3\cos\theta_3
+\frac{2}{3}(\cos\theta_1\sin\theta_2\sin\theta_3\\ \nonumber
&+&\cos\theta_2\sin\theta_1\sin\theta_3
+\cos\theta_3\sin\theta_2\sin\theta_1).
\end{eqnarray}
Which is the same as the $S$ frame result previously obtained in
Eq.~(\ref{cf2}). However when the speed of the boost reaches light
velocity $(\beta\rightarrow1)$, we have:
\begin{eqnarray}
E_{W}(\theta_1\theta_2\theta_3)\rightarrow \tan\theta_1 \tan\theta_2
\tan\theta_3 \sin\Omega_p(2\cos^2\Omega_p - \sin^2\Omega_p),
\end{eqnarray}
where $\sin\Omega_p\rightarrow\sqrt{1-\frac{1}{\Gamma^2}}$ and so,
\begin{eqnarray}
E_{W}(\theta_1\theta_2\theta_3)\rightarrow \tan\theta_1 \tan\theta_2
\tan\theta_3 \sqrt{1-\frac{1}{\Gamma^2}}(\frac{3}{\Gamma^2}-1).
\end{eqnarray}
In the above equation, $\Gamma=\frac{1}{\sqrt{1-\frac{v^2_0}{c^2}}}$
is again the factor of the energy of the particles in the $S$ frame
and $v_0$ is the velocity of the particles in the $S$ frame. We
therefore get for $\theta_i=\theta$ and $\theta'_i=\pi -
\theta=\theta'$:
\begin{eqnarray}
|M|=|M^{\prime}|=|-2E_{W}(\theta_1\theta_2\theta_3)|=\frac{|S_v|}{2}\rightarrow
2\tan^3\theta |\frac{(3-\Gamma^2 )\sqrt{\Gamma^2-1}}{\Gamma^3}|,
\end{eqnarray}
which for $\theta=35.264^{\circ}$ is always less than $2$, showing
non-violation of inequalities regardless of system's energy
($\Gamma$) in the rest frame.

We have therefore shown that when one uses a setup in which $|S_v|$
is maximally violated in the $S$ frame, LT reduces this amount
gradually and that in the extreme relativistic case of
$\beta\rightarrow1$, $|S_v|$ (as well as other) inequalities are
satisfied indicating lack of non-locality. This is consistent with
previous results in the two-particle systems studied in \cite{ahn}.
We have shown this for both W and GHZ states, indicating that this
result is general beyond two-particle systems and regardless of the
type of the entanglement, whether the W or the GHZ state.

Once again, we point to the results of using Pauli's operator
instead of Czachor's in order to investigate the behavior of
$|S_v|$. It has been shown that \cite{mmm}
\begin{eqnarray}\label{amin}
|S_v|&=&\frac{4}{3}|(\cos\Omega_p)^3-\frac{7}{2}(\sin\Omega_p)^2\cos\Omega_p)[(\cos\theta)^3
+ 3\cos\theta-\cos3\theta]|.
\end{eqnarray}
Substituting $\theta=35.264^{\circ}$ into this equation leads to
\begin{eqnarray}\label{awi}
|S_v|&\sim&\vert\frac{19.594}{\Gamma^3}-\frac{15.236}{\Gamma}\vert,
\end{eqnarray}
in the $\beta\rightarrow1$ limit. Indeed, this equation tells us
that $|S_v|$ is not violated by the high energy particles in the
$\beta\rightarrow1$ limit, while, the low energy particles violate
this inequality to the same value as the lab frame whenever the
boost speed reaches light velocity \cite{mmm}. As the GHZ case, we
see that the results of considering Pauli's operator differ from
those obtained by considering Czachor's operator. We next consider
another commonly studied case along the same lines.

\underline{\textbf{Case $II$}}. Here the particles are in the
center-of-mass frame with the following momenta:
\begin{eqnarray}
\overrightarrow{p}_1&=&-p_0\widehat{z} \ \nonumber \\
\overrightarrow{p}_2&=&\frac{p_0}{2}(\widehat{z}+\sqrt{3}\widehat{y})\ \nonumber \\
\overrightarrow{p}_3&=&\frac{p_0}{2}(\widehat{z}-\sqrt{3}\widehat{y}).
\end{eqnarray}
Following Eq.~(\ref{r1}), we find \cite{moradi1}:
\begin{eqnarray}
D(W(\Lambda , p_1))&=&\cos\frac{\Omega_{p}}{2} +i \sigma_y \sin\frac{\Omega_{p}}{2},\\
D(W(\Lambda , p_2))&=&D^\ast (W(\Lambda , p_3))=
\cos\frac{\Omega_{p}}{2}+i(\frac{\sqrt{3}}{2} \sigma_z
-\frac{1}{2}\sigma_y)\sin\frac{\Omega_{p}}{2}.
\end{eqnarray}
In the moving frame, it is a matter of calculation to show that the
GHZ and W states take the form
\begin{eqnarray}
\vert
GHZ^{\Lambda}\rangle&=&-\frac{1}{4\sqrt{2}}\sin^3(\frac{\Omega_p}{2})(\vert
---\rangle - \vert +++\rangle)
\\ \nonumber &+& \cos(\frac{\Omega_p}{2})(\cos^2(\frac{\Omega_p}{2})+\frac{3}{4}\sin^2(\frac{\Omega_p}{2}))\vert GHZ\rangle
\\ \nonumber &+& \frac{1}{\sqrt{2}}[\frac{1}{4}\sin^2(\frac{\Omega_p}{2})\cos(\frac{\Omega_p}{2})(\vert +--\rangle + \vert -++\rangle)
\\ \nonumber &+& \sin(\frac{\Omega_p}{2})(\cos^2(\frac{\Omega_p}{2})+\frac{3}{4}\sin^2(\frac{\Omega_p}{2}))(-\vert -++\rangle +\vert +--\rangle)
\\ \nonumber &+& \frac{1}{2}\sin(\frac{\Omega_p}{2})\cos(\frac{\Omega_p}{2})(\cos(\frac{\Omega_p}{2})+i\frac{\sqrt{3}}{2}\sin(\frac{\Omega_p}{2}))(\vert++-\rangle
-\vert -+-\rangle)\\ \nonumber &+&
\frac{1}{2}\sin^2(\frac{\Omega_p}{2})(\cos(\frac{\Omega_p}{2})-i\frac{\sqrt{3}}{2}\sin(\frac{\Omega_p}{2}))(-\vert--+\rangle
+\vert +-+\rangle)\\ \nonumber
&+&\frac{1}{2}\sin(\frac{\Omega_p}{2})\cos(\frac{\Omega_p}{2})(\cos(\frac{\Omega_p}{2})+i\frac{\sqrt{3}}{2}\sin(\frac{\Omega_p}{2}))(\vert
+-+\rangle +\vert ++-\rangle) ]
\end{eqnarray}
and
\begin{eqnarray}
\vert W\rangle^{\Lambda}&=&\frac{3}{4}\sin^3
(\frac{\Omega_p}{2})\vert +++\rangle-\frac{1}{4}\sin^2
(\frac{\Omega_p}{2})\cos(\frac{\Omega_p}{2})\vert ---\rangle
\\ \nonumber &+& [\cos^3 (\frac{\Omega_p}{2})+\frac{7}{4}\cos(\frac{\Omega_p}{2})\sin^2(\frac{\Omega_p}{2})]\vert -++\rangle
\\ \nonumber &+& [-\sin(\frac{\Omega_p}{2})(\cos(\frac{\Omega_p}{2})+i\frac{\sqrt{3}}{2}\sin(\frac{\Omega_p}{2}))^2+
\frac{1}{4}\sin^3(\frac{\Omega_p}{2})\\ \nonumber
&+&\frac{1}{2}\sin(\frac{\Omega_p}{2})\cos^2
(\frac{\Omega_p}{2})(\cos(\frac{\Omega_p}{2}+i\frac{\sqrt{3}}{2}\sin(\frac{\Omega_p}{2}))]\vert
-+-\rangle\\ \nonumber &+&
[\frac{1}{4}\sin^3(\frac{\Omega_p}{2})-\sin(\frac{\Omega_p}{2})(\cos(\frac{\Omega_p}{2})-i\frac{\sqrt{3}}{2}\sin(\frac{\Omega_p}{2}))^2\\
\nonumber &+&\frac{1}{2}\sin(\frac{\Omega_p}{2})
\cos(\frac{\Omega_p}{2})(\cos(\frac{\Omega_p}{2})-i\frac{\sqrt{3}}{2}\sin(\frac{\Omega_p}{2}))]\vert--+\rangle\\
\nonumber &+& [\frac{1}{2}\sin(\frac{\Omega_p}{2})
\cos(\frac{\Omega_p}{2})(\cos(\frac{\Omega_p}{2})+i\frac{\sqrt{3}}{2}\sin(\frac{\Omega_p}{2}))\\
\nonumber &-&
\frac{1}{2}\sin^2(\frac{\Omega_p}{2})(\cos(\frac{\Omega_p}{2})-i\frac{\sqrt{3}}{2}\sin(\frac{\Omega_p}{2}))+\frac{1}{4}\sin^3(\frac{\Omega_p}{2})]\vert
+--\rangle
\\ \nonumber &+&
[-\frac{1}{4}\sin^2(\frac{\Omega_p}{2})\cos(\frac{\Omega_p}{2})+\cos(\frac{\Omega_p}{2})(\cos(\frac{\Omega_p}{2})-i\frac{\sqrt{3}}{2}\sin(\frac{\Omega_p}{2}))^2
\\ \nonumber &+&\frac{1}{2}\sin^2(\frac{\Omega_p}{2})(\cos(\frac{\Omega_p}{2})-i\frac{\sqrt{3}}{2}\sin(\frac{\Omega_p}{2}))]\vert+-+\rangle
\\ \nonumber &+&
[\cos(\frac{\Omega_p}{2})(\cos(\frac{\Omega_p}{2})+i\frac{\sqrt{3}}{2}\sin(\frac{\Omega_p}{2}))^2-\frac{1}{4}\sin^2(\frac{\Omega_p}{2})\cos(\frac{\Omega_p}{2})
\\ \nonumber &+&\frac{1}{2}\sin^2(\frac{\Omega_p}{2})(\cos(\frac{\Omega_p}{2})+i\frac{\sqrt{3}}{2}\sin(\frac{\Omega_p}{2}))]\vert
++-\rangle.
\end{eqnarray}
Calculations become tedious when one uses Czachor's operator, but
when $\beta \rightarrow 1$, by choosing measurements same as the
\textbf{Case $I$}, we get
\begin{eqnarray}\label{ghzcm}
|M| &\rightarrow & 2\cos\Omega_p(\cos^2\Omega_p + \frac{3}{4}\sin^2\Omega_p)\ \nonumber \\
M& \rightarrow &-M^{\prime}\ , \ \ \ |S_v|=|M+M^{\prime}|
\longrightarrow 0,
\end{eqnarray}
for the $|GHZ\rangle$ state. It is obvious that, independent of the
energy of the particles in the $S$ frame, $|S_v|$ is satisfied.
Using the above equation, one obtains:
\begin{eqnarray}
|M|=|M^{\prime}|\rightarrow\frac{3\Gamma^2 + 1}{2\Gamma^3}\leq 2,
\end{eqnarray}
which unlike \cite{moradi1} shows that $|M|$ and $|M^{\prime}|$ are
satisfied in this limit.

For the $W$ state, when $\beta\rightarrow1$, we obtain
\begin{eqnarray}
|M|=|M^{\prime}|=\frac{|S_v|}{2} \rightarrow \frac{3}{2}\tan^3\theta
(1-\frac{1}{\Gamma^2})^{\frac{3}{2}},
\end{eqnarray}
where $\theta_i=\theta$ has been used as before. We conclude that,
independent of the energy of the particles in the lab frame
($\Gamma$), the inequalities are satisfied when $\beta\rightarrow 1$
and $\theta=35.264^{\circ}$.

As in \textbf{Case $I$}, maximally violated $|S_v|$ setup in the lab
frame leads to the reduction of this violation in the $S^{\prime}$
under LT and leads to lack of non-locality in the extreme
relativistic case of $\beta \rightarrow 1$, regardless of initial
system's energy or the type of entanglement involved.

Additionally, note that two-particle entangled system, in the
center-of-mass frame, has been considered by Ahn et al. \cite{ahn}.
They get:
\begin{eqnarray}
B=\frac{2}{\sqrt{2-\beta^2}}(\cos\Omega_p +\sqrt{1-\beta^2}),
\end{eqnarray}
for Bell's inequality in the moving frame. Clearly, in the lab frame
$(\beta\rightarrow0,\Omega_p\rightarrow0)$, the maximum violation of
the Bell's inequality ($2\sqrt{2}$) is obtained. Note that the
similarity with our results is obvious which indicates that, if the
moving frame uses the same measurements as the lab frame, the degree
of violation decreases under LT leading to lack of entanglement in
the $\beta \rightarrow1$ limit regardless of particles' energy in
the lab frame. These results show that the behavior of the
three-particle non-local system under LT is the same as the
two-particle entangled system, if we use $|S_v|$ and proper
measurements which violate $|S_v|$ to its maximum possible value in
the $S$ frame. This is important to note that our results indicate
that the general behavior of entanglement in the Bell, GHZ and W
states under LT is the same as each other, if one uses Czachor's
operator, the Bell inequality for the Bell states \cite{ahn}, and
the special set of measurements violating $|S_v|$ to its maximum
possible violation amount in the $S$ frame for the GHZ and W states.

Finally, we setout to obtain the behavior of $|S_v|$ for both the
GHZ and W states, when Pauli's operator is used instead of Czachor's
operator. In this situation, we get
\begin{eqnarray}
E_{GHZ}(ABC)&=&a\cos(\phi_{1}+\phi_{2}+\phi_{3})+b
\cos(\phi_{1}-\phi_{2}-\phi_{3})+i(\sin(\phi_{1}+\phi_{2}-\phi_{3})\nonumber \\
\nonumber &+&\sin(-\phi_{1}+\phi_{2}-\phi_{3}))f
-2\cos(\phi_{1}-\phi_{2}+\phi_{3})d\\
&+&(-\cos(\phi_{1}-\phi_{2}+\phi_{3})+\cos(\phi_{1}+\phi_{2}-\phi_{3}))e,
\end{eqnarray}
where $i=\sqrt{-1}$ and
\begin{eqnarray}
a&=&-\frac{1}{16}\sin^6(\frac{\Omega_p}{2})+\cos^2(\frac{\Omega_p}{2})
(\cos^2(\frac{\Omega_p}{2})+\frac{3}{4}\sin^2(\frac{\Omega_p}{2}))^2,\\
\nonumber
b&=&\frac{1}{16}\sin^4(\frac{\Omega_p}{2})\cos^2(\frac{\Omega_p}{2})+
\sin^2(\frac{\Omega_p}{2})(\cos^2(\frac{\Omega_p}{2})+\frac{3}{4}\sin^2(\frac{\Omega_p}{2}))^2,\\
\nonumber
f&=&\frac{1}{4}\sin^2(\frac{\Omega_p}{2})\cos^2(\frac{\Omega_p}{2})
(\cos^2(\frac{\Omega_p}{2})+\frac{3}{4}\sin^2(\frac{\Omega_p}{2}))^2,\\
\nonumber
d&=&\frac{1}{4}\sin^3(\frac{\Omega_p}{2})\cos(\frac{\Omega_p}{2})
(\cos^2(\frac{\Omega_p}{2})+\frac{3}{4}\sin^2(\frac{\Omega_p}{2})),\\
\nonumber
e&=&\frac{1}{4}\sin^4(\frac{\Omega_p}{2})(\cos^2(\frac{\Omega_p}{2})
+\frac{3}{4}\sin^2(\frac{\Omega_p}{2})).
\end{eqnarray}
Now, setting $\phi_i=\frac{\pi}{4}$ and
$\phi_i^{\prime}=\frac{3\pi}{4}$, after some calculations we get
\begin{eqnarray}
E_{GHZ}(ABC)&=&-\frac{\sqrt{2}}{2}a+\frac{\sqrt{2}}{2}b-\sqrt{2}d=-E_{GHZ}(A'B'C'),\\
\nonumber
E_{GHZ}(A'BC)&=&-\frac{\sqrt{2}}{2}a+\frac{\sqrt{2}}{2}b+\sqrt{2}d,\\
\nonumber
E_{GHZ}(AB'C)&=&-\frac{\sqrt{2}}{2}a-\frac{\sqrt{2}}{2}b+i\sqrt{2}f-\sqrt{2}d,\\
\nonumber
E_{GHZ}(ABC')&=&-\frac{\sqrt{2}}{2}a-\frac{\sqrt{2}}{2}b+\sqrt{2}d,\\
\nonumber
E_{GHZ}(A'B'C)&=&+\frac{\sqrt{2}}{2}a+\frac{\sqrt{2}}{2}b-i\sqrt{2}f-\sqrt{2}d-\sqrt{2}e,\\
\nonumber
E_{GHZ}(AB'C')&=&+\frac{\sqrt{2}}{2}a-\frac{\sqrt{2}}{2}b-\sqrt{2}d,\\
\nonumber
E_{GHZ}(A'BC')&=&+\frac{\sqrt{2}}{2}a-\frac{\sqrt{2}}{2}b+\sqrt{2}d+\sqrt{2}e,
\end{eqnarray}
which finally leads to:
\begin{eqnarray}\label{svet1}
\vert S_v \vert=\sqrt{32a^2 +8f^2},
\end{eqnarray}
which, as a check, covers the lab frame result ($4\sqrt{2}$) in the
appropriate limit $\Omega_p\rightarrow0$. Since in the
$\beta\rightarrow1$ limit, $f\rightarrow0$ for the low energy
particles, the $|S_v|$ inequality is violated to the same value as
the lab frame ($4\sqrt{2}$), which is in agreement with previous
study pointed out in the Case $I$ \cite{mmm}. It can also be checked
that, in the $\beta\rightarrow1$ limit, high energy particles does
not violate $|S_v|$. In fact, in this situation we get
$|S_v|\approx2.121496$. This behavior is in line with some studies
on two-particle systems \cite{terashima,terashima1}.

For the W state, calculations lead to
\begin{eqnarray}
E_W(\theta_1 \theta_2 \theta_3)
&=&\cos(\theta_1)\cos(\theta_2)\cos(\theta_3)A'+
\sin(\theta_1)\sin(\theta_2)\sin(\theta_3)B' \nonumber \\
\nonumber &+& \sin(\theta_1)\cos(\theta_2)\cos(\theta_3)C'+
\sin(\theta_1)\cos(\theta_2)\sin(\theta_3)D'\\
\nonumber &+& \sin(\theta_1)\sin(\theta_2)\cos(\theta_3)E'+
\cos(\theta_1)\sin(\theta_2)\sin(\theta_3)F'\\
&+& \cos(\theta_1)\sin(\theta_2)\cos(\theta_3)G'+
\cos(\theta_1)\cos(\theta_2)\sin(\theta_3)H',
\end{eqnarray}
in which
\begin{eqnarray}
A'&=&a'^2 -b'^2-c'^2 +d'^2 +e'^2 - g'^2 -h'^2 +f'^2\\
\nonumber B'&=& 2a'b'+2c'f'+2d'g'+2e'h'\\
\nonumber C'&=& 2a'c'+2b'e'-2g'e'-2h'd'\\
\nonumber D'&=& 2a'd'+2c'h'-2b'g'-2e'f'\\
\nonumber E'&=& 2a'e'+2g'c'-2b'h'-2d'f'\\
\nonumber F'&=& 2a'f'+2g'h'-2b'c'-2d'e'\\
\nonumber G'&=& 2a'g'+2b'd'-2c'e'-2f'h'\\
\nonumber H'&=& 2a'h'+2b'e'-2c'd'-2f'g',
\end{eqnarray}
and
\begin{eqnarray}
a'&=&\frac{\sqrt{3}}{4}\sin^3(\frac{\Omega_p}{2}) \\ \nonumber
b'&=&-\frac{1}{4\sqrt{3}}\sin^2(\frac{\Omega_p}{2})\cos(\frac{\Omega_p}{2})\\
\nonumber
c'&=&\frac{1}{\sqrt{3}}[\cos^3(\frac{\Omega_p}{2})+\frac{7}{4}\cos(\frac{\Omega_p}{2})\sin^2(\frac{\Omega_p}{2})]\\
\nonumber
d'&=&\frac{1}{\sqrt{3}}[\sin(\frac{\Omega_p}{2})(\cos(\frac{\Omega_p}{2})+i\frac{\sqrt{3}}{2}\sin(\frac{\Omega_p}{2}))^2
+\frac{1}{4}\sin^3(\frac{\Omega_p}{2})\\
\nonumber &+& \frac{1}{2}\sin(\frac{\Omega_p}{2})\cos^2(\frac{\Omega_p}{2})(\cos(\frac{\Omega_p}{2})+i\frac{\sqrt{3}}{2}\sin(\frac{\Omega_p}{2}))]\\
\nonumber
e'&=&\frac{1}{\sqrt{3}}[\frac{1}{4}\sin^3(\frac{\Omega_p}{2})-\sin(\frac{\Omega_p}{2})(\cos(\frac{\Omega_p}{2})-i\frac{\sqrt{3}}{2}\sin(\frac{\Omega_p}{2}))^2
\\
\nonumber &+&\frac{1}{2}\sin(\frac{\Omega_p}{2})\cos(\frac{\Omega_p}{2})(\cos(\frac{\Omega_p}{2})-i\frac{\sqrt{3}}{2}\sin(\frac{\Omega_p}{2}))]\\
\nonumber
f'&=&\frac{1}{\sqrt{3}}[\frac{1}{2}\sin(\frac{\Omega_p}{2})\cos(\frac{\Omega_p}{2})(\cos(\frac{\Omega_p}{2})+i\frac{\sqrt{3}}{2}\sin(\frac{\Omega_p}{2}))
+\frac{1}{4}\sin^3(\frac{\Omega_p}{2})\\
\nonumber &-&\frac{1}{2}\sin^2(\frac{\Omega_p}{2})(\cos(\frac{\Omega_p}{2})-i\frac{\sqrt{3}}{2}\sin(\frac{\Omega_p}{2}))]\\
\nonumber
g'&=&\frac{1}{\sqrt{3}}[-\frac{1}{4}\sin^2(\frac{\Omega_p}{2})\cos(\frac{\Omega_p}{2})+\cos(\frac{\Omega_p}{2})(\cos(\frac{\Omega_p}{2})-i\frac{\sqrt{3}}{2}\sin(\frac{\Omega_p}{2}))^2
\\
\nonumber &+&\frac{1}{2}\sin^2(\frac{\Omega_p}{2})(\cos(\frac{\Omega_p}{2})-i\frac{\sqrt{3}}{2}\sin(\frac{\Omega_p}{2}))]\\
\nonumber
h'&=&\frac{1}{\sqrt{3}}[\cos(\frac{\Omega_p}{2})(\cos(\frac{\Omega_p}{2})+i\frac{\sqrt{3}}{2}\sin(\frac{\Omega_p}{2}))^2-\frac{1}{4}\sin^2(\frac{\Omega_p}{2})\cos(\frac{\Omega_p}{2})
\\
\nonumber
&+&\frac{1}{2}\sin^2(\frac{\Omega_p}{2})(\cos(\frac{\Omega_p}{2})+i\frac{\sqrt{3}}{2}\sin(\frac{\Omega_p}{2}))].
\end{eqnarray}
Now, for the correlation functions we get
\begin{eqnarray}
&E&_W(\theta_1 \theta_2 \theta_3)=0.5443 A'+0.1924 B'+0.38(C'+G'+H')+0.272(D'+E'+F')\\
\nonumber &E&_W(\theta'_1 \theta'_2 \theta'_3)=-0.5443 A'+0.1924 B'+0.38(C'+G'+H')-0.272(D'+E'+F')\\
\nonumber &E&_W(\theta'_1 \theta_2 \theta_3)=-0.5443 A'+0.1924 B'+0.38(C'-G'-H')+0.272(D'+E'-F')\\
\nonumber &E&_W(\theta_1 \theta'_2 \theta_3)=-0.5443 A'+0.1924 B'+0.38(-C'+G'-H')+0.272(-D'+E'+F')\\
\nonumber &E&_W(\theta_1 \theta_2 \theta'_3)=-0.5443 A'+0.1924 B'+0.38(-C'-G'+H')+0.272(D'-E'+F')\\
\nonumber &E&_W(\theta'_1 \theta'_2 \theta_3)=0.5443 A'+0.1924 B'+0.38(-C'-G'+H')+0.272(-D'+E'-F')\\
\nonumber &E&_W(\theta'_1 \theta_2 \theta'_3)=0.5443 A'+0.1924 B'+0.38(-C'+G'-H')+0.272(D'-E'-F')\\
\nonumber &E&_W(\theta_1 \theta'_2 \theta'_3)=0.5443 A'+0.1924
B'+0.38(C'-G'-H')+0.272(-D'-E'+F'),
\end{eqnarray}
which finally gives
\begin{eqnarray}
|S_v|=|-2.1772A'+1.088(D'+E'+F')|.
\end{eqnarray}
One may check to see that the result of lab frame ($4.354$) is
recovered in the appropriate limit $\Omega_p=0$. In addition, it is
a matter of calculation to show that, in the $\beta\rightarrow1$
limit, this inequality is preserved by the high energy particles
($\Gamma\rightarrow\infty$). Moreover, for the low energy particles
($\Gamma\rightarrow1$) in the $\beta\rightarrow1$ limit, the
violation amount of $|S_v|$ is the same as the lab frame ($4.354$).
Therefore, we can conclude that if one uses Pauli's operator, Bell's
inequality for the Bell states \cite{terashima,terashima1}, and
special set of measurements violating $|S_v|$ to its maximum
violation amount in the lab frame, then the violation of the
corresponding Bell inequality in the moving frame depends on the
particles energy in the lab frame.

\section{Summary and Concluding Remarks \label{Conclusions}}
In this work we have considered various three-particle systems and
have calculated the Bell-like inequalities for each systems in the
moving frame ($S^{\prime}$) under Lorentz transformation. Our main
result is that Svetlichny's inequality ($|S_v|$) when combined with
Czachor's relativistic spin operator gives consistent and reasonable
results in line with various studies in two-particle systems
\cite{terashima,terashima1}, devoid of inconsistencies when one uses
other measures such as $|M|$ or $|M^{\prime}|$. We also studied the
results of using Pauli's operator to investigate the behavior of
$|S_v|$ under LT. The results of considering Pauli's operator is in
agreement with some previous studies on the two and three-particle
systems \cite{terashima,terashima1,you,mmm} and differ from those
obtained by considering Czachor's operator.

We are able to show that, whenever Czachor's operator is considered,
if one uses the same set of angles in $S$ as well as $S^{\prime}$
and starts with such a setup that maximizes non-locality in the rest
frame then the results are such that $(\textmd{i})$ non-locality
decreases as a function of boost parameter $\beta$. $(\textmd{ii})$
in the extreme relativistic case of $\beta \rightarrow 1$ limit, the
inequality is satisfied indicating lack of non-locality.
$(\textmd{iii})$ in such a limit, though the value of the inequality
itself depends on the initial particles' energy, it is not violated
no matter how that energy is chosen. $(\textmd{iv})$ all these
results are true regardless of the type of entanglement present in
the pure three-particle system: the GHZ or W state. $(\textmd{v})$
these results are true regardless of how one sets up the particles
in the rest frame, net non-zero momentum (\textbf{Case $I$}) or
center-of-mass frame (\textbf{Case $II$}). Furthermore, comparison
of our results with that of Ref.~\cite{moradi1}, where a different
set of measurements are used to evaluate $|M|$, indicates that the
three-particle Bell-like inequalities are more sensitive to
measurement set-ups than the bi-partite case of Bell's inequality, a
point has been emphasized in previous non-relativistic studies
\cite{cerc,AA}.

The fact that our general results are consistent with previous
studies in two-particle systems
\cite{alsing,gingrich,li,rem,caban,pacho,ahn,terashima,terashima1,lee,kim},
supports these results against some other results to their contrary
\cite{moradi1,moradi2,you,mmm}. However, it is clear that one can
also find certain measurements that leave the correlation functions
unchanged in the moving frame, thus leading to invariance of such
inequalities \cite{alsing,terashima,terashima1,lee,kim,moradi2,you}
as well as finding particular cases which lead to maximization of
their violation in certain moving frames \cite{jordan,cafaro}.
Eventually, our results show that the inconsistency between the
previous attempts mentioned in the introduction is the direct result
of using the $|M|$ inequality and the set of measurements which
violate the $|M|$ inequality to its maximum violation amount in the
lab frame to study the behavior of the GHZ state under LT
\cite{moradi1,you}. In fact, if one uses Czachor's operator and the
special set of measurements violating the $|S_v|$ inequality to its
maximum possible violation amount in the lab frame the mentioned
inconsistency will be eliminated independently of considering the
$|S_v|$, $|M|$ or $|M'|$ inequality.

We also investigated the results of using Pauli's operator instead
of Czachor's operator to study the behavior of $|S_v|$. Our study
shows that in this situation the violation amount of $|S_v|$ in the
moving frame depends on the particles energy in the lab frame. This
result is in line with some previous studies
\cite{terashima,terashima1,you,mmm} and also helps in eliminating
the mentioned inconsistency. Finally, we note that since our results
for Pauli's operator vs. Czachor's operator lead to decidedly
different type of behavior under LT, they provide a mechanism
whereby a Stern-Gerlach experiment could be used to see which
results are more consistent with experiments and therefore provide
evidence for a more suitable spin operator.
%%%%%%%%%%%%%%%%%%%%%%%%%%%%%%%%%%%%%%%%%%%%%%%%%%%%%%%%%
\section*{Acknowledgements}
The work of H. Moradpour has been supported financially by Research
Institute for Astronomy and Astrophysics of Maragha (RIAAM) under
research No.$1/3720-77$. The work of A. Montakhab is supported by
Shiraz University Research Council.
%%%%%%%%%%%%%%%%%%%%%%%%%%%%%%%%%%%%%%%%%%%%%%%%%%%%%%%%%%%%

\end{document}